\documentclass[12pt]{iopart}
\usepackage[dvips]{graphicx}
\begin{document}
\title{Efficient charge pumping in graphene}
\author{B. Abdollahipour$^{1}$ and R. Mohammadkhani$^{2}$
}
\address{$^1$ Faculty of Physics, University of Tabriz, Tabriz
51666-16471, Iran\\
$^2$ Department of Physics, Faculty of Science,
University of Zanjan, Zanjan 45371-38791, Iran}
\begin{abstract}
We investigate a graphene quantum pump, adiabatically driven by two
thin potential barriers vibrating around their equilibrium
positions. For the highly doped leads, the pumped current per mode
diverges at the Dirac point due to the more efficient contribution
of the evanescent modes in the pumping process. The pumped current
shows an oscillatory behavior with an increasing amplitude as a
function of the carrier concentration. This effect is in contrast to
the decreasing oscillatory behavior of the similar normal pump. The
graphene pump driven by two vibrating thin barriers operates more
efficient than the graphene pump driven by two oscillating thin
barriers.
\end{abstract}
\pacs{73.23.-b, 72.80.Vp, 73.40.Gk}
\maketitle

\section{Introduction}

Quantum pumping is a coherent transport mechanism to produce a DC
charge current in the absence of an external bias voltage by an
appropriate periodical variation of the system parameters. The idea
of quantum pumping was introduced by Thouless, who proposed a
quantum pump driven by a moving potential for a DC current
generation \cite{Thouless83}. Then, Brouwer developed an elegant
formula for the adiabatic quantum pumping in an open quantum system
based on the scattering matrix approach \cite{Brouwer98}. Finally,
Moskalets and B\"{u}ttiker generalized the scattering matrix
approach for the Ac transport and they also derived general
expressions of the pumped current, heat flow, and shot noise for an
adiabatically driven quantum pump in the weak pumping limit
\cite{Moskalets02}. It has been shown that the pumping current is
related to the geometric (Berry) phases \cite{Makhlin01} and quantum
interference effects \cite{Zhou99}. Several proposals have been
proposed application of the quantum pumping as a potential way for
generation of a dissipationless charge current in the nanoscale
devices \cite{Altshuler99} as well as a promising method for
generating a dynamically controlled flow of spin-entangled electrons
\cite{Das06}, a way to produce a spin polarized current which has a
main importance in the spintronics \cite{Avishai10} and a method to
transfer charge in a quantized fashion \cite{Wang13}. Quantum
pumping has been realized experimentally in different nanoscale
systems such as, quantum dots with Coulomb blockade
\cite{Switkes99,Watson03,Buitelaar08} and Josephson junctions
\cite{Vartiainen07,Giazotto11}.

Recently, experimental realization of graphene, a monolayer of
carbon atoms with hexagonal lattice structure, has introduced a new
type of two dimensional materials with unique properties
\cite{Novoselov04,Novoselov05,Zhang05}. Electrons in graphene behave
identical to two dimensional massless Dirac fermions due to the
gapless semiconducting electronic band structure with a linear
dispersion relation at low energies \cite{Neto09,Sarma11}. Such a
peculiar quasiparticle spectrum accompanied by the unique feature of
Chirality has led to anomalous behaviors in several transport
phenomena including the Klein tunneling \cite{Katsnelson06}, minimum
of the conductivity \cite{Tworzydlo06}, integer quantum Hall effect
\cite{Gusynin05} and Josephson effect \cite{Beenakker08}.

Several graphene quantum pumps have been proposed and different
aspects of them have been investigated as a result of the unique
properties of graphene. Prada {\it et al.} have proposed a graphene
pump driven by two oscillating square potential barriers. Their
study revealed that the evanescent modes in graphene have a dominant
contribution in the pumped current which gives rise to a universal
dimensionless pumping efficiency at the Dirac point \cite{Prada09}.
It has also been shown that adding stationary magnetic barriers in
the graphene pump leads to a valley-polarized and pure valley pumped
currents \cite{Grichuk13}. Zhu and Chen have studied a quantum pump
device composed of a ballistic graphene coupled to the reservoirs
via two oscillating tunnel barriers \cite{Zhu09}. Bercioux {\it et
al.} have shown that presence of a gate tunable spin-orbit
interaction can generate a spin-polarized pumped current
\cite{Bercioux12}. It has been shown that combination of high
frequency vibrations and metallic transport in graphene makes it
extremely suitable for charge pumping due to the sensitivity of its
transport coefficients to perturbations in electrostatic potential
and mechanical deformations \cite{Low12}. Tiwari and Blaauboer have
found out that combination of a perpendicular magnetic field in the
central pumping region with two oscillating electrical voltages in
the leads causes both charge and spin pumped currents through
traveling modes \cite{Tiwari10}. It has been shown that presence of
a superconducting lead enhances the pumped current per mode by a
factor of 4 at a resonance condition \cite{Alos-Palop11}. Kundu {\it
et al.} have shown that a graphene superconducting double barrier
structure supports large values of pumped charge when the pumping
contour encloses a resonance point \cite{Kundu11}. The effect of the
interlayer coupling on the pumped current in a bilayer graphene pump
has been investigated by Wakker {\it et al.} \cite{Wakker10}. A
large pumped current around the Dirac point has been demonstrated in
the bipolar regime by a single-parameter graphene pump invoking
graphene's intrinsic features of chirality and bipolarity
\cite{San-Jose11}.

In this paper, we introduce a mechanism for efficient charge pumping
in graphene. We propose a graphene quantum pump driven by two thin
potential barriers vibrating around their equilibrium positions. We
analyze the pumped current in two different cases of the highly
doped and undoped leads. We compare the results with the more
familiar case of the charge pumping by the oscillating thin
barriers. The results of the later case is very similar to the
graphene pump considered in Ref. \cite{Prada09}. We find a very
efficient pumping by the vibrating thin barriers in comparison to
the oscillating ones. In the case of highly doped leads and for
vibrating thin barriers pumped current per mode diverges at the
Dirac point, whereas for the oscillating thin barriers it tends to a
limited value. An interesting and distinguished feature of the
pumped current generated by the vibrating thin barriers is its
increasing oscillatory behavior as a function of the carrier
concentration. Decreasing oscillatory behavior of the similar normal
pump reveal that it is a unique feature of the Dirac fermions pump
and then, it can be attributed to the linear dispersion relation of
the electrons in graphene. This feature is in contrast to the
tendency of the pumped current generated by the oscillating thin
barriers to a limited value at high carriers concentration.

The outline of the paper is as follows. In the section \ref{Model}
we introduce the proposed pump and basic equations which are used
for calculation of the pumped current. Section \ref{Highly} is
devoted to study the pumped current in the case of the highly doped
leads. In the section \ref{Undoped} we give the results and
discussions for the pump with the undoped leads. Section
\ref{Features} is devoted to a discussion about the main features of
the pump and its experimental implementation. Finally, we summarize
our results in the section \ref{Conclusion}.

\section{Model and basic equations}\label{Model}

\begin{figure}
\centerline{\includegraphics[width=7cm]{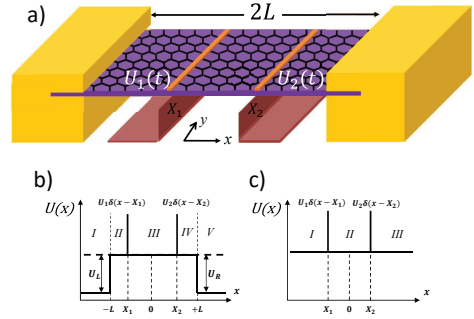}} \caption{(Color online) a) Graphene quantum pump driven by
two thin barriers imposed via two thin gates. The periodic variations of the magnitudes or positions of the
thin barriers are considered as pumping parameters. b) Electrostatic potential through the system in the case
of highly doped leads and (c) undoped leads. The solid lines show the instantaneous potential profile during
the pumping cycle.} \label{system-figure}
\end{figure}

We consider a quantum pump composed of a graphene sheet with length
$2L$ and width $W$ connected to two leads kept at zero bias voltage.
It is driven out of equilibrium by two thin potential barriers as it
has been shown in Fig. (\ref{system-figure}). These thin barriers
can be realized by electric fields or thin gates under the graphene.
We study two cases of the highly doped and undoped leads and compare
their results. In the case of highly doped leads evanescent modes
are induced in the pumping region. It is in contrast to the case of
the undoped leads which all of the modes in the pumping region are
propagating. This feature allows us to investigate the contribution
of the evanescent modes in the pumped current \cite{Prada09}. In
order to adiabatically pump a charge current between two leads kept
at zero bias voltage, the scattering properties of the system should
undergo a slow and periodic variation. It is achieved by cyclic
changing some parameters of the system usually referred as pumping
parameters. The slow variation is attained when the pumping
parameters vary slower than the dwell time of the carriers in the
pump region. In this work, we consider two different methods to
drive the pump. These methods are realized by two oscillating or
vibrating thin barriers. In the first case, the pump is driven by
two thin barriers located at the fixed positions $X_1$ and $X_2$
with the oscillating magnitudes of the potentials
\begin{eqnarray}\label{Pumping-parameters-U}
\nonumber&&U_1(t)=U_{1,0}+\delta U_1\cos(\omega t),
\\&& U_2(t)=U_{2,0}+\delta
U_2\cos(\omega t+\varphi),
\end{eqnarray}
where $U_{1,0}$ and $U_{2,0}$ are static potentials, $\delta U_1$
and $\delta U_2$ are amplitudes of the oscillations and $\varphi$ is
the phase difference of them. This driving method is very similar to
a previous work \cite{Prada09}. The second way, is a mechanism
usually referred as the "snow plow" mechanism \cite{Avron00}. Pump
is driven by two thin barriers with fixed magnitudes of the
potentials and periodic variation of their positions as it is given
below
\begin{eqnarray}\label{Pumping-parameters-X}
\nonumber&&X_1(t)=X_{1,0}+\delta X_1\cos(\omega t),
\\&& X_2(t)=X_{2,0}+\delta
X_2\cos(\omega t+\varphi),
\end{eqnarray}
where $X_{1,0}$ and $X_{2,0}$ are equilibrium positions of the thin
barriers, $\delta X_1$ and $\delta X_2$ are amplitudes of the
vibrations and $\varphi$ is their phase difference. If we denote the
two pumping parameters by $\eta_1$ and $\eta_2$, the adiabatic
pumped current in the bilinear response regime, where
$\delta\eta_1\ll\eta_1$ and $\delta\eta_2\ll\eta_2$, is given by
\cite{Brouwer98,Prada09}
\begin{eqnarray}\label{Brouwer-formula}
I_P=I_0\sum_{\alpha=L,R}Im\left(\frac{\partial
S_{L,\alpha}}{\partial \eta_1}\frac{\partial
S^{\ast}_{L,\alpha}}{\partial \eta_2}\right),
\end{eqnarray}
where $I_0=(\omega/2\pi)e\delta \eta_1\delta \eta_2\sin\varphi$ and
$S_{L,\alpha}$ is an element of the scattering matrix. In the above
equation summation goes over the transverse modes in the left and
right leads. The pumped charge depends only on the area spanned by
the pumping cycle in the parameters space and not on its details.
This equation shows that pumped current in the adiabatic limit is
proportional to the variation frequency and vanishes in the zero
phase difference, when the area inclosed in the parameters space is
zero during a cycle.

To calculate the pumping current we need to obtain the scattering
matrix of the pump. The low energy excitations in the graphene are
described by the two dimensional Dirac equation
\begin{equation}\label{Dirac-equation}
\left[\upsilon_{F}\mathbf{p}\cdot\hat{\mbox{\boldmath
$\sigma$}}+U(x)\right]\Psi=E\Psi,
\end{equation}
where $\mathbf{p}$ is the momentum operator relative to the Dirac
point, $\hat{\mbox{\boldmath $\sigma$}}=(\sigma_x,\sigma_y)$ is the
vector of the Pauli matrices and $U(x)$ is the potential energy
across the system. We model the thin barriers by symmetric delta
function potentials in our calculations. Thus, in the pumping region
$U(x)=U_1\delta(x-X_1)+U_2\delta(x-X_2)$ and in the leads
$U_{L,R}\rightarrow -\infty$ for highly doped leads and $U_{L,R}=0$
in the case of undoped leads. In fact, the highly doped leads model
normal metal leads connected to graphene. Figs.
(\ref{system-figure}b) and (\ref{system-figure}c) show the potential
profiles through the system in the cases of highly doped and undoped
leads, respectively. $\Psi$ in the Eq. (\ref{Dirac-equation}) is a
two component spinor in the pseudospin space which refers to the two
sublattices in the two dimensional honeycomb lattice. We solve Eq.
(\ref{Dirac-equation}) in different regions of the pump in the cases
of the highly doped and undoped leads in the following sections. Due
to the conservation of the transverse momentum through the system,
mode matching gives us the elements of the scattering matrix.
%
%
\section{Highly doped leads}\label{Highly}

As it has been shown in the Fig. (\ref{system-figure}b), system has
five regions in the case of highly doped leads. In the leads, where
$U_{L,R}\rightarrow -\infty$, carriers densities are very large in
contrast to the pumping region. This situation is realized by the
metallic leads. The highly doped leads induce evanescent modes in
the pumping region and it leads to the contribution of the
evanescent modes in the pumped current. The wave functions in the
left ($x<-L$) and right ($x>L$) leads are,
\begin{eqnarray}
\Psi_{L}=e^{i(K_L x+qy)}\left(
\begin{array}{cc}
\begin{array}{c}
1
\end{array}
\\*
\begin{array}{c}
1
\end{array}
\end{array}
\right)+ r e^{i(-K_Lx+qy)}\left(
\begin{array}{cc}
\begin{array}{c}
1
\end{array}
\\*
\begin{array}{c}
-1
\end{array}
\end{array}
\right),
\end{eqnarray}
\begin{eqnarray}
\Psi_{R}=te^{i(K_Rx+qy)}\left(
\begin{array}{cc}
\begin{array}{c}
1
\end{array}
\\*
\begin{array}{c}
1
\end{array}
\end{array}
\right),
\end{eqnarray}
where $r$ and $t$ are reflection and transmission coefficients,
respectively. In the above relations $q$ is the transverse momentum
and  $K_{L,R}=\sqrt{\left(\frac{E-U_{L,R}}{\hbar
v_F}\right)^{2}-q^{2}}$ are the wave vectors in the leads. In the
pumping region, region between the left lead and first delta
potential $-L<x<X_1$, region between two delta potentials
$X_1<x<X_2$ and region between second delta potential and the right
lead $X_2<x<L$ which are denoted by $j=1,2,3$ respectively, wave
functions read
\begin{eqnarray}
\Psi_{j}=a_{j}e^{i(kx+qy)}\left(
\begin{array}{cc}
\begin{array}{c}
1
\end{array}
\\*
\begin{array}{c}
e^{i\phi}
\end{array}
\end{array}
\right)+ b_{j}e^{i(-kx+qy)}\left(
\begin{array}{cc}
\begin{array}{c}
1
\end{array}
\\*
\begin{array}{c}
-e^{-i\phi}
\end{array}
\end{array}
\right),
\end{eqnarray}
where $k=\sqrt{\left(\frac{E}{\hbar v_F}\right)^{2}-q^{2}}$ is the
wave vector in the pumping region and $\phi=\tan^{-1}(\frac{q}{k})$
is the incident angle. The boundary conditions for these wave
functions are the continuity at the $x=\pm L$ and satisfying the
following conditions at the positions of the delta potentials
\begin{eqnarray}
\Psi_{2}|_{x=X_1}=\mathcal{T}_1\Psi_{1}|_{x=X_1},~
\Psi_{3}|_{x=X_2}=\mathcal{T}_2\Psi_{2}|_{x=X_2},
\end{eqnarray}
where
\begin{eqnarray}\label{transfer-matrix}
\mathcal{T}_{1,2}=\frac{1+i\sigma_xU_{1,2}/2\hbar
v_F}{1-i\sigma_xU_{1,2}/2\hbar
v_F}=e^{2i\sigma_x\tan^{-1}(\frac{U_{1,2}}{2\hbar v_F})}.
\end{eqnarray}
This boundary condition is obtained by integrating the Dirac
equation through a symmetric delta function potential
\cite{Titov07}. We obtain transmission and reflection coefficients
by solving these equations. The obtained expressions are very
lengthy to be given here. In the following we calculate the pumped
current in the case of the highly doped leads for oscillating and
vibrating thin barriers.
%
\begin{figure}
\centerline{\includegraphics[width=7cm]{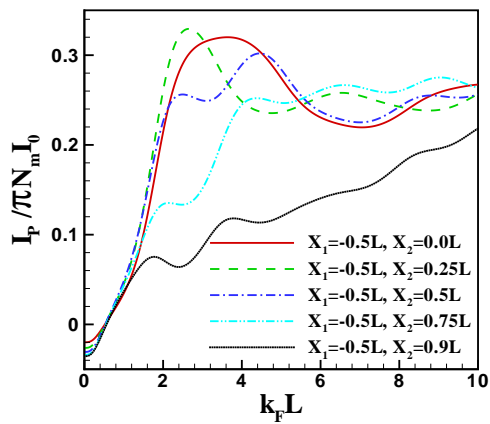}} \caption{(Color online) Normalized pumped current as a
function of the carrier concentration, $k_FL$, for different configurations of the oscillating thin barriers.
Where we have considered $U_{1,0}=U_{2,0}=0$.} \label{Ip-highly-doped-u}
\end{figure}
%
%
\begin{figure}
\centerline{\includegraphics[width=7cm]{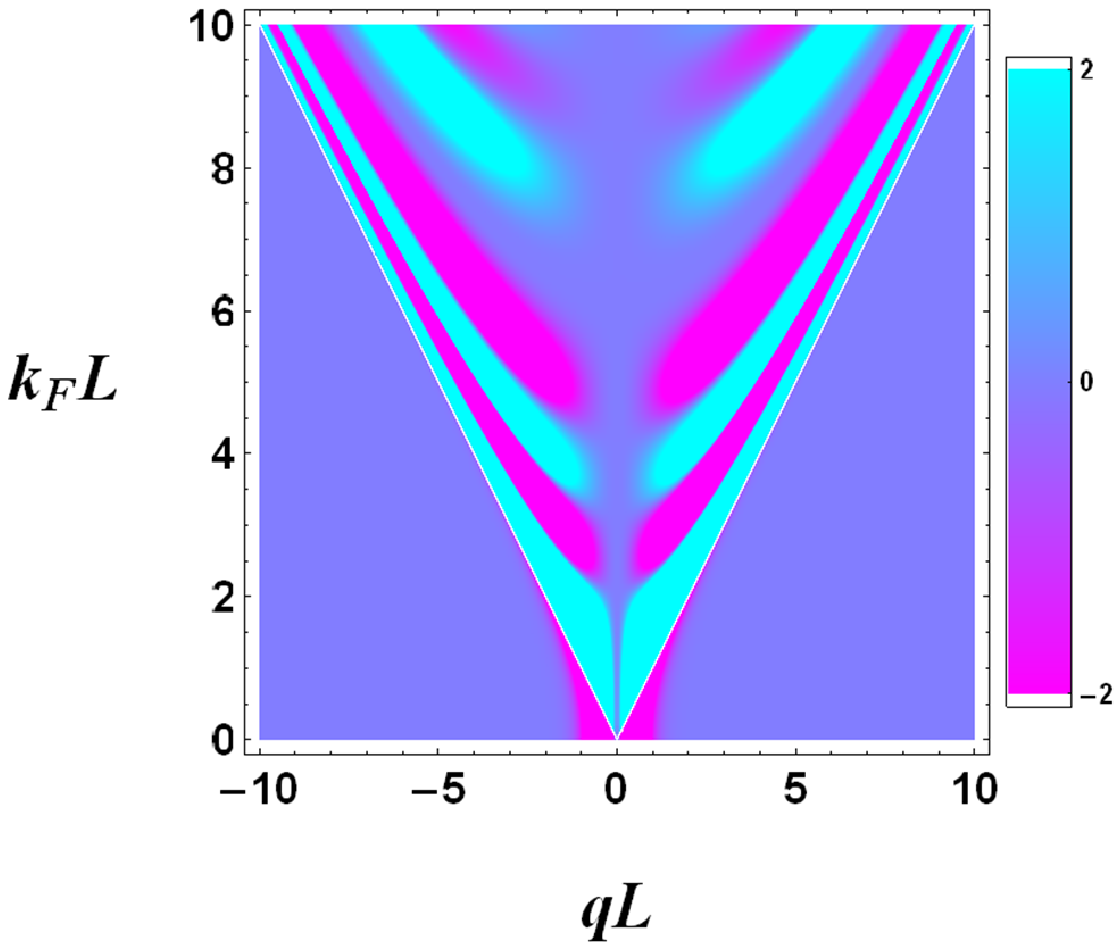}} \caption{(Color online) Momentum distribution of the
normalized pumped current as a function of the transverse momentum $q$ and the carrier concentration $k_FL$ in
the case of the oscillating thin barriers for $U_{1,0}=U_{2,0}=0$ and $X_2=-X_1=0.5L$. The white lines
indicate the points with $|q|=k_F$. Around the Dirac point, the evanescent modes ($|q|>k_F$) contribute in the
pumped current in opposite direction of the extended modes ($|q|\leq k_F$).} \label{MD-HD-OS}
\end{figure}
%
\subsection{Driving by the oscillating thin barriers}

Let us focus to the pumped current generated by variations of the
magnitudes of two thin potential barriers given by Eq.
(\ref{Pumping-parameters-U}). In this case the pumped current is
obtained by the following relation
\begin{eqnarray}\label{Pumping-current-1}
I_P=I_0\sum_{n}Im\left\{\frac{\partial r}{\partial
U_1}\frac{\partial r^{\ast}}{\partial U_2}+\frac{\partial
t}{\partial U_1}\frac{\partial t^{\ast}}{\partial U_2}\right\},
\end{eqnarray}
where $I_0=(\omega/2\pi)e\delta U_1\delta U_2\sin\varphi$ and
summation is over the transverse modes denoted by $n$. For short and
wide graphene ($W\gg L$) we can change the summation over $n$ to
integration over the continuous transverse momentum,
$\sum_q\rightarrow (W/2\pi)\int dq$. Thus, the pumping current reads
\begin{eqnarray}\label{Pumping-current-2}
I_P=N_mI_0\int^{+\infty}_{-\infty}\frac{dq}{k_F}Im\left\{\frac{\partial
r}{\partial U_1}\frac{\partial r^{\ast}}{\partial
U_2}+\frac{\partial t}{\partial U_1}\frac{\partial
t^{\ast}}{\partial U_2}\right\},
\end{eqnarray}
where $N_m=4k_FW/\pi$ is the number of the propagating modes in the
pump and $k_F$ is the Fermi momentum. The coefficient $4$ is due to
the degeneracy including two valleys and two spin states in
graphene. In Fig. (\ref{Ip-highly-doped-u}) we have shown the
normalized pumped current $I_P/\pi N_m I_0$, as a function of $k_FL$
(characterizing the carrier concentration) for different
configurations of the potential barriers. These plots reveal some
important features. First, the pumped current changes sign around
$k_FL\sim 1/2$. It happens due to the generation of the pumped
current by the evanescent modes ($|q|>k_F$) in opposite direction of
the pumped current generated by the extended modes ($|q|\leq k_F$),
around the Dirac point. To explain it, we have shown the kernel or
the momentum distribution of the normalized pumped current in the
Fig. \ref{MD-HD-OS}. It shows that around the Dirac point
oscillating thin barriers drive electrons occupying extended and
evanescent modes in different directions. At a specific value of
$k_FL$ these opposite contributions cancel each other and the pumped
current vanishes. This feature can help to distinguish between the
pumped and the rectified currents \cite{Brouwer01}. Second, minimum
of the pumped current, pumped current at the Dirac point, depends on
the configuration of the potential barriers and in contrast to the
Ref. \cite{Prada09}, it has not an universal value. Third, there is
a weak oscillatory behavior in the curves which is caused by the
quantum interferences due to the reflections from two thin potential
barriers. Fourth, all of the plots tend to the same value $I_P/\pi
N_m I_0$=0.25 in the limit of large $k_FL$. It happens irrespective
of the pump configuration and it is identical to the result of Ref.
\cite{Prada09}.
%
\begin{figure}
\centerline{\includegraphics[width=7cm]{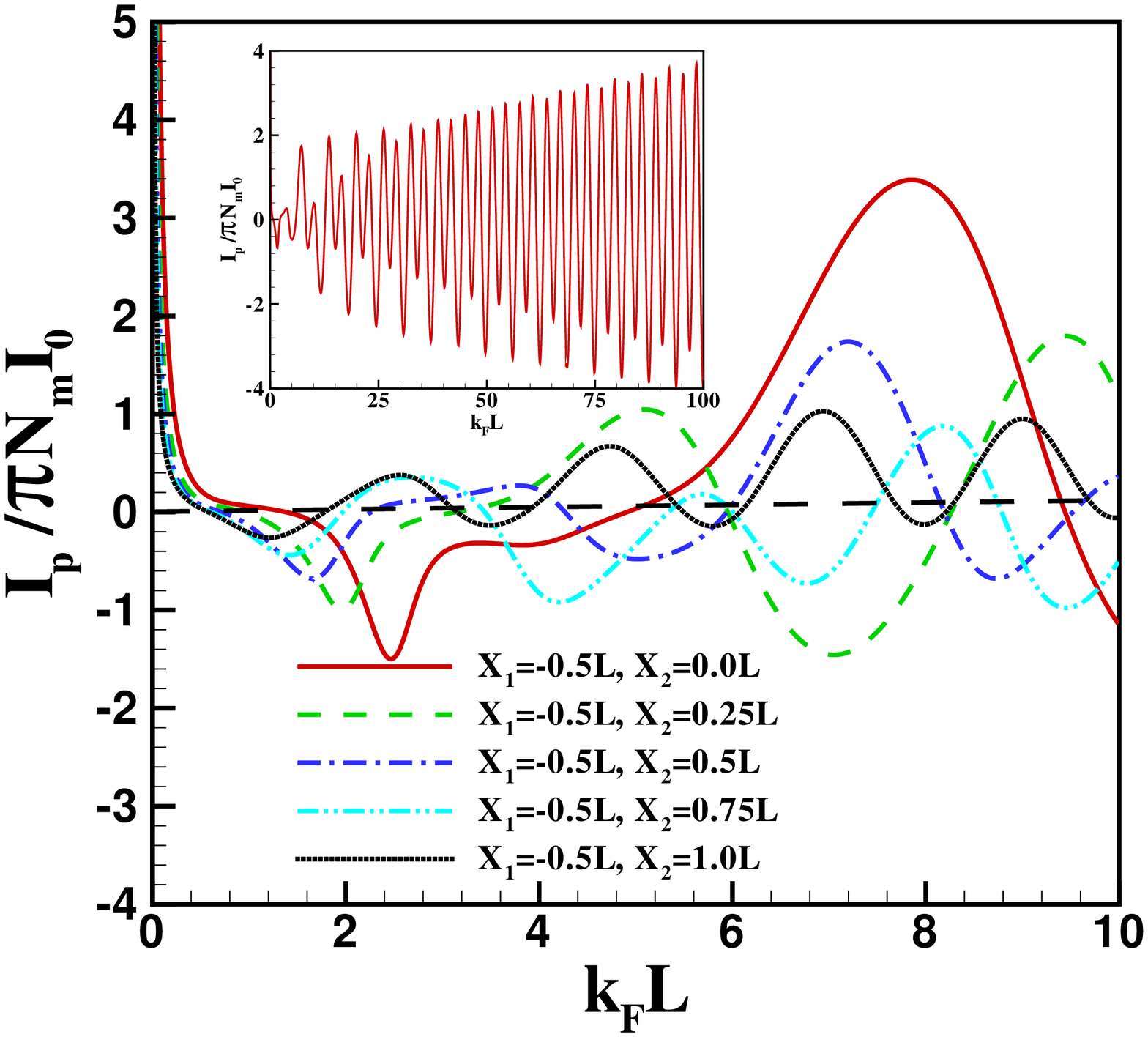}} \caption{(Color online) Normalized pumped current as a
function of the carrier concentration $k_FL$, for different configurations of the vibrating thin barriers.
Where $U_{1}=U_{2}=\hbar v_F$ has been considered. Inset: Normalized pumped current as a function of the
carrier concentration in the wide range of $k_FL$ for $X_2=-X_1=0.5L$.} \label{Ip-highly-doped-X}
\end{figure}
%
\subsection{Driving by the vibrating thin barriers}
%
In this section we consider the pumped current generated by the
vibration of two thin potential barriers around their equilibrium
positions, given by Eq. (\ref{Pumping-parameters-X}). For this case
the pumped current reads
\begin{eqnarray}\label{Pumping-current-3}
I_P=
N_mI_0\int^{+\infty}_{-\infty}\frac{dq}{k_F}Im\left\{\frac{\partial
r}{\partial X_1}\frac{\partial r^{\ast}}{\partial
X_2}+\frac{\partial t}{\partial X_1}\frac{\partial
t^{\ast}}{\partial X_2}\right\},
\end{eqnarray}
where $I_0=(\omega/2\pi)e\delta X_1\delta X_2\sin\varphi$. In the
calculations we consider that $U_{1}=U_{2}=\hbar v_F$. Fig.
(\ref{Ip-highly-doped-X}) shows the normalized pumped current as a
function of $k_FL$. As it is apparent from the figure, there are
main differences between the pumped current generated by the
position and magnitude variations of the potential barriers. In the
case of driving by position variation, the normalized pumped
current, pumped current per mode normalized by $I_0$, diverges at
the Dirac point. It indicates the nonzero value of the pumped
current at the vanishingly small density of states around the Dirac
point. The pumped current shows asymmetric oscillations around the
zero as a function of the carrier concentration. For vibrating thin
barriers the pumped current keeps an increasing oscillatory behavior
by increasing $k_FL$, as it has been shown in the inset of Fig.
(\ref{Ip-highly-doped-X}). It is in contrast with the oscillating
thin barriers which the pumped current tends to a limited value at
large $k_FL$.

\begin{figure}
\centerline{\includegraphics[width=7cm]{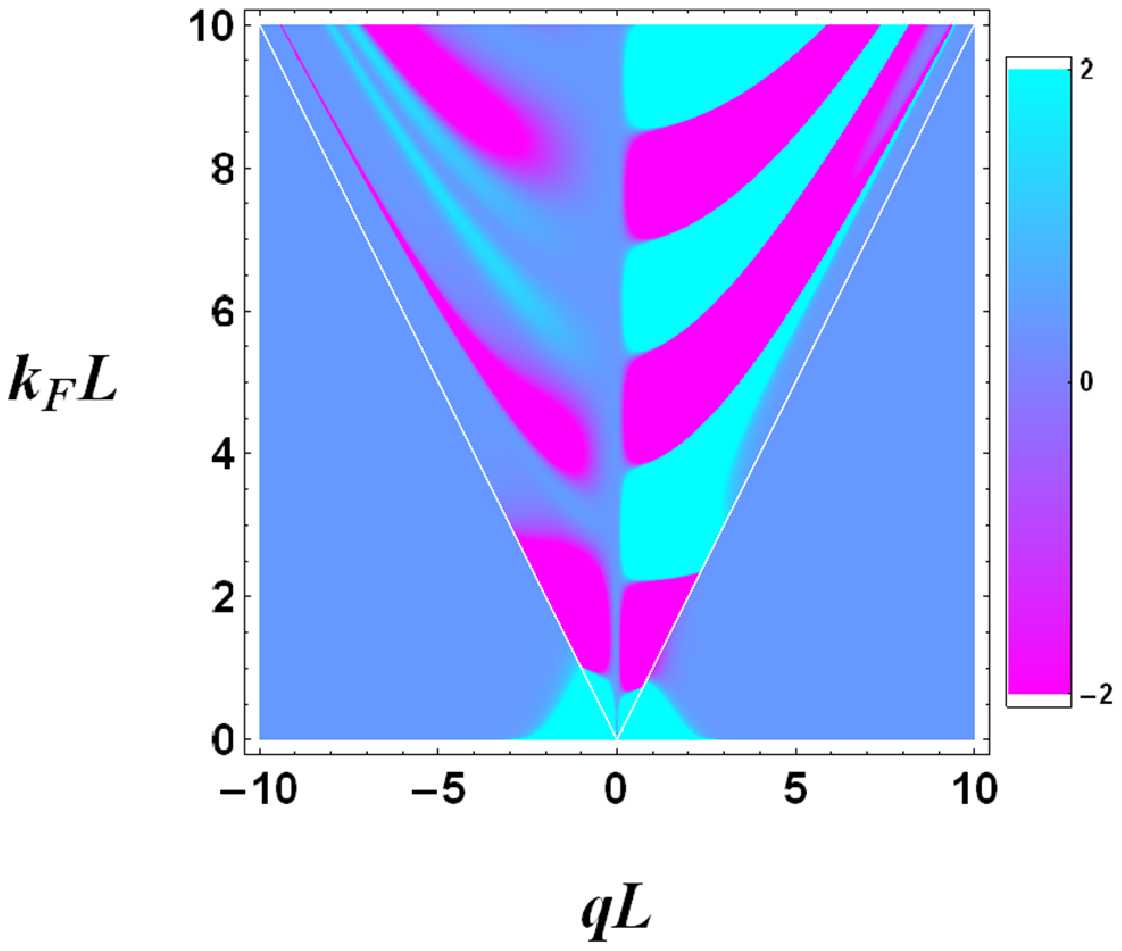}} \caption{(Color online) Comparison of the momentum
distributions of the normalized pumped currents as a function of the transverse momentum $q$ and the carrier
concentration $k_FL$ for the oscillating (left half) and the vibrating (right half) thin barriers. Here we
have considered $U_{1,0}=U_{2,0}=\hbar v_F$ and $X_{2,0}=-X_{1,0}=0.5L$. The white lines indicate the points
with $|q|=k_F$.} \label{MD-HD}
\end{figure}

To clarify the obtained results, we compare the momentum
distribution of the pumped current for the oscillating and vibrating
thin barriers. The momentum distribution is a symmetric function of
the transverse momentum $q$. Thus, it lets us to plot the momentum
distributions for two different driving methods in one figure. Left
half of Fig. (\ref{MD-HD}) shows momentum distribution for the
oscillating thin barriers and the right half of it belongs to the
vibrating thin barriers for $U_{1,0}=U_{2,0}=\hbar v_F$ and
$X_{2,0}=-X_{1,0}=0.5L$. As it is apparent in the figure, in both
cases the contribution of the normal incident Dirac fermions ($q=0$)
in the pumped current is zero due to the Klein tunneling.
Contribution of the evanescent modes in the pumped current around
the Dirac point is considerable in both cases. In spite of these
similarities there is a main difference between two methods of
driving. In the case of the vibrating thin barriers the contribution
of the extended modes increases by increasing carrier concentration,
whereas it decreases in the case of the oscillating thin barriers.
This feature makes pumping by the vibrating thin barriers more
effective than the oscillating ones.
%
%
%
\section{Undoped leads}\label{Undoped}

There are three different regions in the system with undoped leads.
The region on the left of the first delta potential $x<X_1$, region
between two delta potentials $X_1<x<X_2$ and the region on the right
of the second delta potential $X_2<x$. The wave functions in the
left and right leads are as follows
\begin{eqnarray}
\Psi_{L}=e^{i(kx+qy)}\left(
\begin{array}{cc}
\begin{array}{c}
1
\end{array}
\\*
\begin{array}{c}
e^{i\phi}
\end{array}
\end{array}
\right)+ r e^{i(-kx+qy)}\left(
\begin{array}{cc}
\begin{array}{c}
1
\end{array}
\\*
\begin{array}{c}
-e^{-i\phi}
\end{array}
\end{array}
\right),
\end{eqnarray}
\begin{eqnarray}
\Psi_{R}=te^{i(kx+qy)}\left(
\begin{array}{cc}
\begin{array}{c}
1
\end{array}
\\*
\begin{array}{c}
e^{i\phi}
\end{array}
\end{array}
\right).
\end{eqnarray}
In the pump region the wave function is
\begin{eqnarray}
\Psi_{p}=ae^{i(kx+qy)}\left(
\begin{array}{cc}
\begin{array}{c}
1
\end{array}
\\*
\begin{array}{c}
e^{i\phi}
\end{array}
\end{array}
\right)+ be^{i(-kx+qy)}\left(
\begin{array}{cc}
\begin{array}{c}
1
\end{array}
\\*
\begin{array}{c}
-e^{-i\phi}
\end{array}
\end{array}
\right).
\end{eqnarray}
where $k=\sqrt{\left(\frac{E}{\hbar v_F}\right)^{2}-q^{2}}$ is the
wave vector in the pumping region as well as leads and
$\phi=\tan^{-1}(\frac{q}{k})$ is the incident angle. These wave
functions should satisfy the following boundary conditions
\begin{eqnarray}
\Psi_{p}|_{x=X_1}=\mathcal{T}_1\Psi_{L}|_{x=X_1},~\Psi_{R}|_{x=X_2}=\mathcal{T}_2\Psi_{p}|_{x=X_2}.
\end{eqnarray}
$\mathcal{T}_{1,2}$ are given by the Eq. (\ref{transfer-matrix}).
Solution of these equations yields the reflection and transmission
coefficients. Since, the obtained expressions are lengthy we will
not give them here.
%
\begin{figure}
\centerline{\includegraphics[width=7cm]{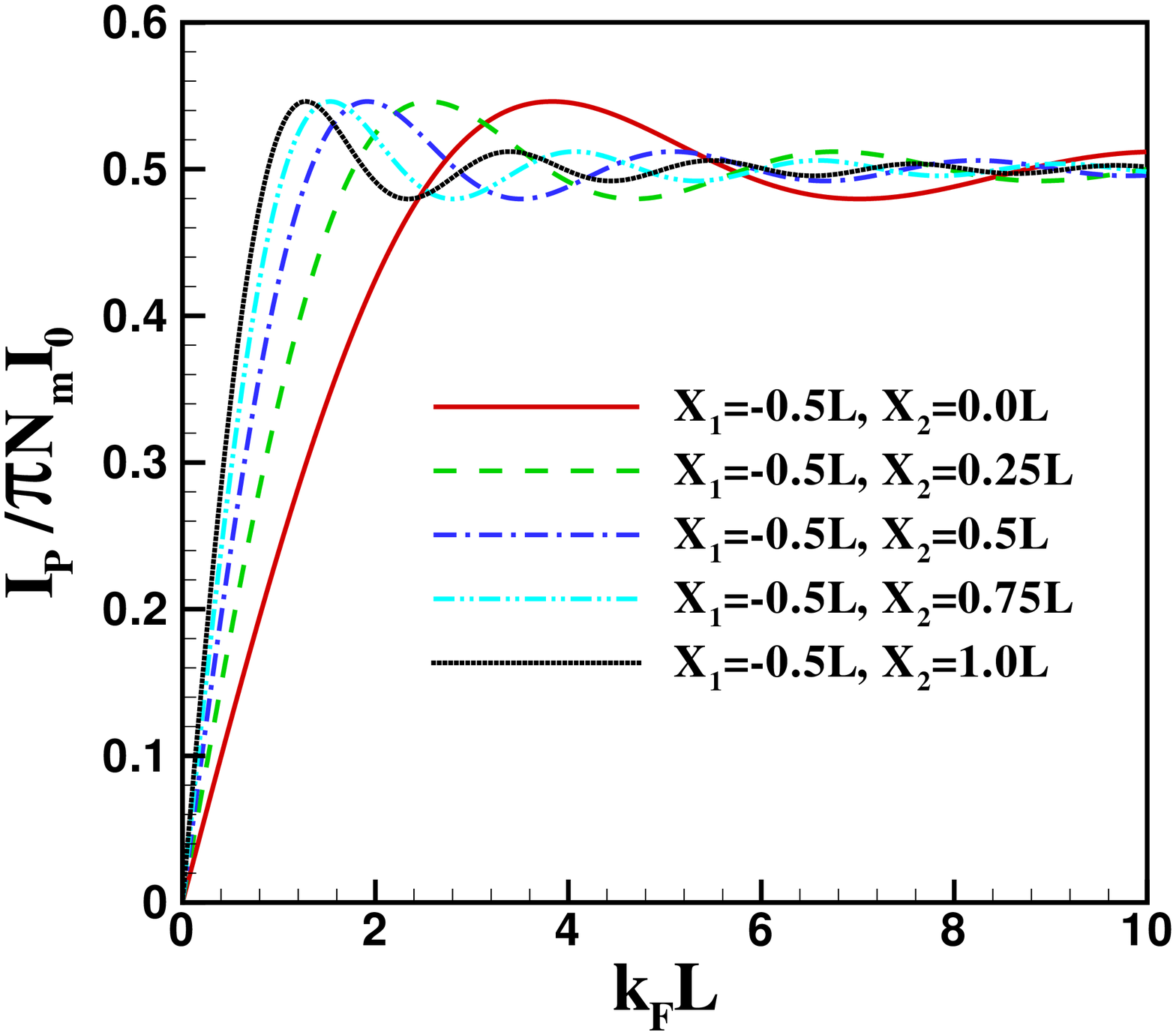}} \caption{(Color online) Normalized pumped current as a
function of the carrier concentration, $k_FL$, for different configurations of the oscillating thin barriers.
Here $U_{1,0}=U_{2,0}=0$.} \label{Ip-undoped-U}
\end{figure}
%
\subsection{Oscillating thin barriers}
%
In this section we present the results for the oscillating thin
barriers in the case of the undoped leads. The pumped current is
given by Eq. (\ref{Pumping-current-2}) and we obtain a simple
analytical equation for it,
\begin{eqnarray}\label{Pumping-current-5}
I_P=N_mI_0\int^{+\infty}_{-\infty}\frac{dq}{k_F}\frac{q^2}{k^2}\sin\left(2k(X_2-X_1)\right).
\end{eqnarray}
As it has been shown in Fig. (\ref{Ip-undoped-U}), the normalized
pumped current vanishes at the Dirac point due to the absence of the
extended and evanescent modes. It has an oscillatory behavior as a
function of the carrier concentration arisen by quantum
interferences. At large values of $k_FL$ the normalized pumped
current tends to a limited value $I_P/\pi N_mI_0=0.5$, as also
indicated in the Ref. \cite{Prada09}.
%
\subsection{Vibrating thin barriers}
\begin{figure}
\centerline{\includegraphics[width=7cm]{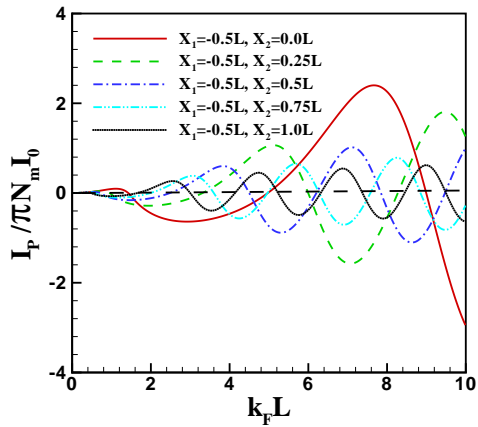}} \caption{(Color online) Normalized pumped current as a
function of the carrier concentration $k_FL$, for different configurations of the vibrating thin barriers.
Here $U_{1}=U_{2}=\hbar v_F$.} \label{Ip-undoped-X}
\end{figure}

%
For the vibrating thin barriers the pumped current is given by Eq.
(\ref{Pumping-current-3}). The normalized pumped current has been
shown in Fig. (\ref{Ip-undoped-X}) for different configurations of
the pump and $U_{1}=U_{2}=\hbar v_F$. As it is apparent from Fig.
(\ref{Ip-undoped-X}), the pumped current vanishes around the Dirac
point. At the Dirac point all of the modes in the pump are
unpopulated and there is not nonzero contribution in the pumped
current. It is due to the absence of the extended modes at the Dirac
point and the evanescent modes in the graphene connected to the
undoped leads. As in the case of the highly doped leads, there is an
increasing oscillatory behavior in the normalized pumped current as
a function of $k_FL$.
%
%
%
\section{Unique features}\label{Features}

\begin{figure}
\centerline{\includegraphics[width=7cm]{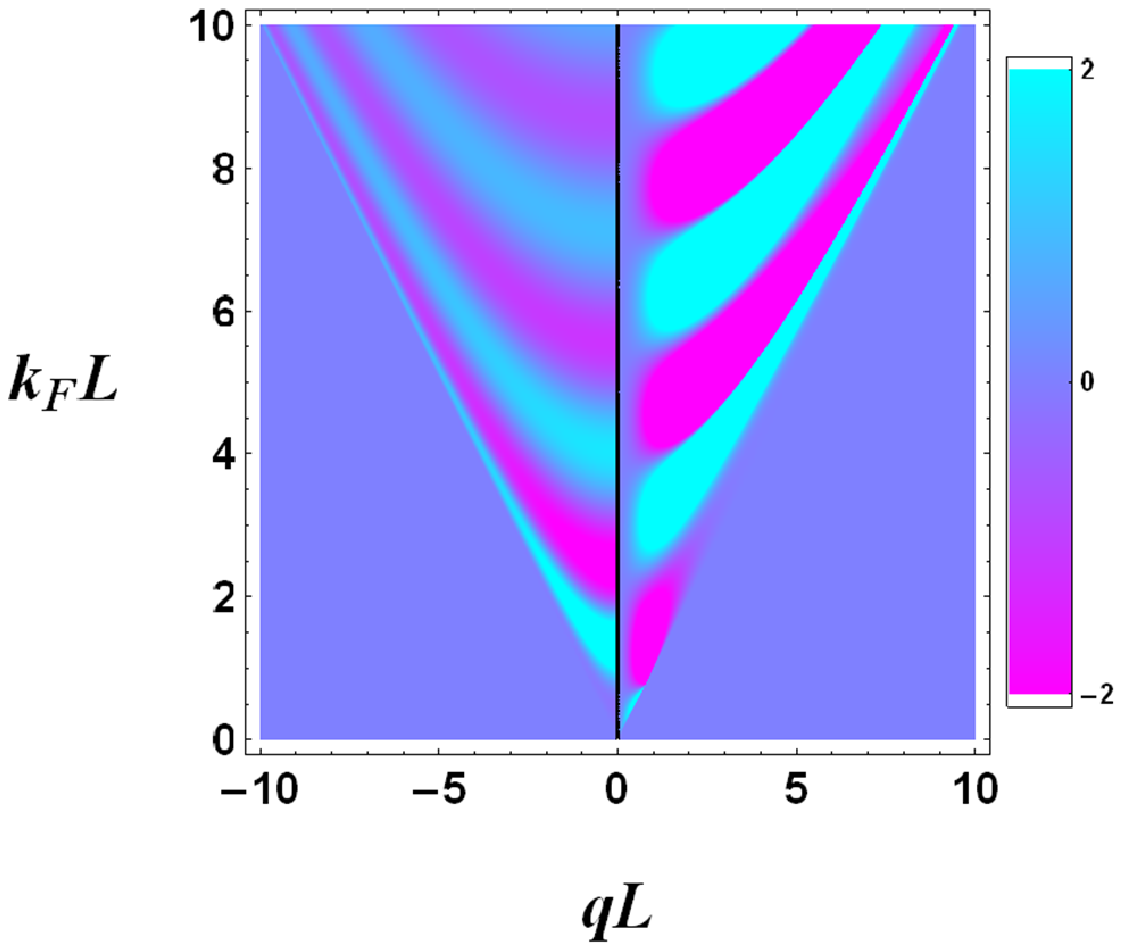}} \caption{(Color online) Comparison of the momentum
distributions of the normalized pumped currents as a function of the transverse momentum $q$ and the carrier
concentration $k_FL$ for the normal (left half) and graphene (right half) pumps driven by vibrating thin
barriers. Here we have considered $U_{1}=U_{2}=\hbar v_F$ and $X_2=-X_1=0.5L$.} \label{MD-UD-NG}
\end{figure}

For emphasizing the unique feature of the pumping by vibrating thin
barriers in graphene, we compare it with a similar normal pump (a
similar pump based on the 2DEG). We notice that, the pumped current
in graphene shows similar behaviors for large values of the carrier
concentration in both cases of the highly doped and the undoped
leads. Thus, we just compare the results in the case of the undoped
leads. For the normal pump we should solve the Schrodinger equation
in the presence of the two delta function potentials and then, mode
matching gives us the reflection and the transmission coefficients.
As like as graphene, the pumped current for the normal pump shows an
oscillatory behavior as a function of $k_FL$. But, in spite of the
graphene its amplitude decreases by increasing carrier
concentration. To be clear, we compare the momentum distributions of
the normal and graphene pumps. In Fig. (\ref{MD-UD-NG}), the left
half shows momentum distribution for the normal pump and the right
half belongs to the graphene pump. The oscillatory behavior in both
cases is clearly apparent in the figure. But, contribution of the
extended modes in the pumped current follows opposite directions by
increasing the carrier concentration. It increases in the graphene
pump, whereas it decreases in the normal pump. Thus, we can conclude
that the increasing contribution of the extended modes in the pumped
current is a unique feature for driving of the Dirac fermions by the
vibrating potential barriers.

Let us to discuss the practical situations. In the case of the
highly doped leads the normalized pumped current ($i_p=I_P/\pi
N_mI_0 $) generated by the oscillating thin barriers tends to a
limited value at the Dirac point. It means that at $k_FL\rightarrow
0$, where the number of the extended modes $N_m$ vanishes, the
pumped current $I_P$ should also vanish. On the other hand, the
normalized pumped current generated by the vibrating thin barriers
diverges at the dirac point. But, when $k_FL\rightarrow 0$ and
$i_p\rightarrow\infty$, $k_FLi_p\sim 1$ and the pumped current tends
to a nonzero value at the Dirac point. we can estimate the magnitude
of the pumped current by considering an adiabatic pumping frequency
in the range of the terahertz, $\omega/2\pi\sim 1.0$ THz
\cite{Prada09}. Assuming typical values in the experiments for other
parameters, $W/L\sim 10-100$, $\delta X_{1,2}/L\sim 0.01-0.1$ and
$\varphi=\pi/2$, we have a pumped current around $I_P\sim 0.5-500$
nA. It is due to the efficient contribution of the evanescent modes
in the pumped current generated by the vibrating thin barriers. This
value for the pumped current at the Dirac point is well accessible
in the experiment.
%
%
%
\section{Conclusion}\label{Conclusion}
%
In conclusion we have investigated a new mechanism for driving the
graphene pump. In this graphene pump two thin potential barriers are
employed to drive the system out of equilibrium via two different
methods of magnitude oscillation and position vibration. The case of
driving by oscillating thin barriers has a similarity to the
graphene pump considered in the Ref. \cite{Prada09}. As it is
expected, at the large carrier concentrations our results tend to
their results. It is due to the fact that, at large carrier
concentrations the exact configuration of the pump has an
insignificant effect in the pumped current. But, there are important
differences in vicinity of the Dirac point. The minimum of the
pumped current, arising due to the contribution of the evanescent
modes, depends on the pump configuration and it has not a universal
value in spite of the Ref. \cite{Prada09}. Also, the pumped current
changes sign around the Dirac point due to the opposite
contributions of the evanescent and extended modes. On the other
hand, new features appear in the case of the vibrating thin
barriers. The pumped current has an increasing oscillatory behavior
around zero as a function of the carrier concentration. It is due to
the increasing contribution of the extended modes in the pumped
current by increasing the carrier concentration. Comparison with the
similar normal pump indicates that it is a unique feature of the
Dirac fermions. Also, the normalized pumped current diverges at the
Dirac point due to the more effective contribution of the evanescent
modes when the thin barriers have nonzero magnitudes. Due to these
facts, we can conclude that driving by oscillating and vibrating
thin barriers are very effective methods to generate a pumped
current in graphene. Thus, despite of the practical difficulties for
experimental realization of the considered pump we believe that it
has a more chance to be confirmed in the experiment.

we have considered the thin potential barriers as a delta function
potentials in our calculations. It is a limited situation and in
practice we expect a determinate width for the thin potential
barriers. It results to a complex dependence of the pumped current
on the width and height of the potential barriers. We will consider
this situation in the subsequent works.


%

\section*{References}

\end{document}